\documentclass[aps,prl, twocolumn,, reprint, superscriptaddress]{revtex4-1}
\usepackage{graphicx}
\usepackage{amsmath}
\usepackage{wasysym}
\usepackage{mathrsfs}
\usepackage{color}
\usepackage{bm}
\usepackage{amssymb}
\usepackage{hyperref}

\usepackage{enumerate}
\newcommand{\beq}{\begin{equation}}
\newcommand{\eeq}{\end{equation}}
\newcommand{\bea}{\begin{eqnarray}}
\newcommand{\eea}{\end{eqnarray}}

\newcommand\gD{D}

\def\ket#1{{\left|#1\right\rangle}}

\begin{document}
\draft
\title{Order by Disorder and by Doping 
 in Quantum Hall Valley Ferromagnets}
\author{Akshay Kumar}
\email{akfour@princeton.edu}
\affiliation{Department of Physics, Princeton
University, Princeton, NJ 08544, USA}
\affiliation{Max-Planck-Institut f\"{u}r Physik Komplexer Systeme, Dresden 01187, Germany}
\author{S.  A. Parameswaran}
\email{sidp@uci.edu}
\affiliation{Department of Physics and Astronomy, University of California, Irvine, CA 92697, USA}
\author{S. L. Sondhi}
\email{sondhi@princeton.edu}
\affiliation{Department of Physics, Princeton
University, Princeton, NJ 08544, USA}
\affiliation{Max-Planck-Institut f\"{u}r Physik Komplexer Systeme, Dresden 01187, Germany}

\begin{abstract}
We examine the Si(111) multi-valley quantum Hall system and show that it exhibits an exceptionally rich interplay of broken symmetries and 
quantum Hall ordering already near integer fillings $\nu$ in the range $\nu=0-6$. This six-valley system has a large 
$[SU(2)]^3\rtimes D_3$  symmetry in the limit where the magnetic length is much larger than the lattice constant. We find that the discrete ${D}_3$ factor breaks over a broad range of fillings at a finite temperature transition to a discrete nematic phase. As $T \rightarrow 0$ the $[SU(2)]^3$ continuous symmetry also breaks: completely near $\nu =3$, to a residual $[U(1)]^2\times SU(2)$ near $\nu=2$ and $4$ and to a residual $U(1)\times [SU(2)]^2$ near $\nu=1$ and $5$. Interestingly, the symmetry breaking near $\nu=2,4$
and $\nu=3$ involves a combination of selection by thermal fluctuations known as ``order by disorder'' and a selection by the
energetics of Skyrme lattices induced by moving away from the commensurate fillings, a mechanism we term ``order by doping''.
We also exhibit modestly simpler analogs in the four-valley Si(110) system.
\end{abstract}
\maketitle

\noindent\textbf{Introduction:} 
Even with the recent
 explosive increase in the number of interesting condensed matter systems such as
topological insulators and ultracold atoms, the venerable quantum Hall effect (QHE) continues to hold its own, 
owing to its exceptional tunability coupled with the
unique physics of QH states. The latter allows different forms
of topological order to exist under a potentially infinite set of conditions involving different 
degrees of commensuration between particle and magnetic flux density, captured by the filling factor $\nu$. The former allows several energy scales to be varied
largely independently and --- most
 relevant  for this paper ---  admits the possibility of engineering multi-component QH
systems which can then exhibit an interesting interplay between broken symmetries and topological order---as in the phenomenon of quantum Hall ferromagnetism~\cite{das_sarma_qh, sondhi_skyrmions_1993, Moon_etal}.

\begin{figure}[tb]
\centering
\includegraphics[width=\columnwidth]{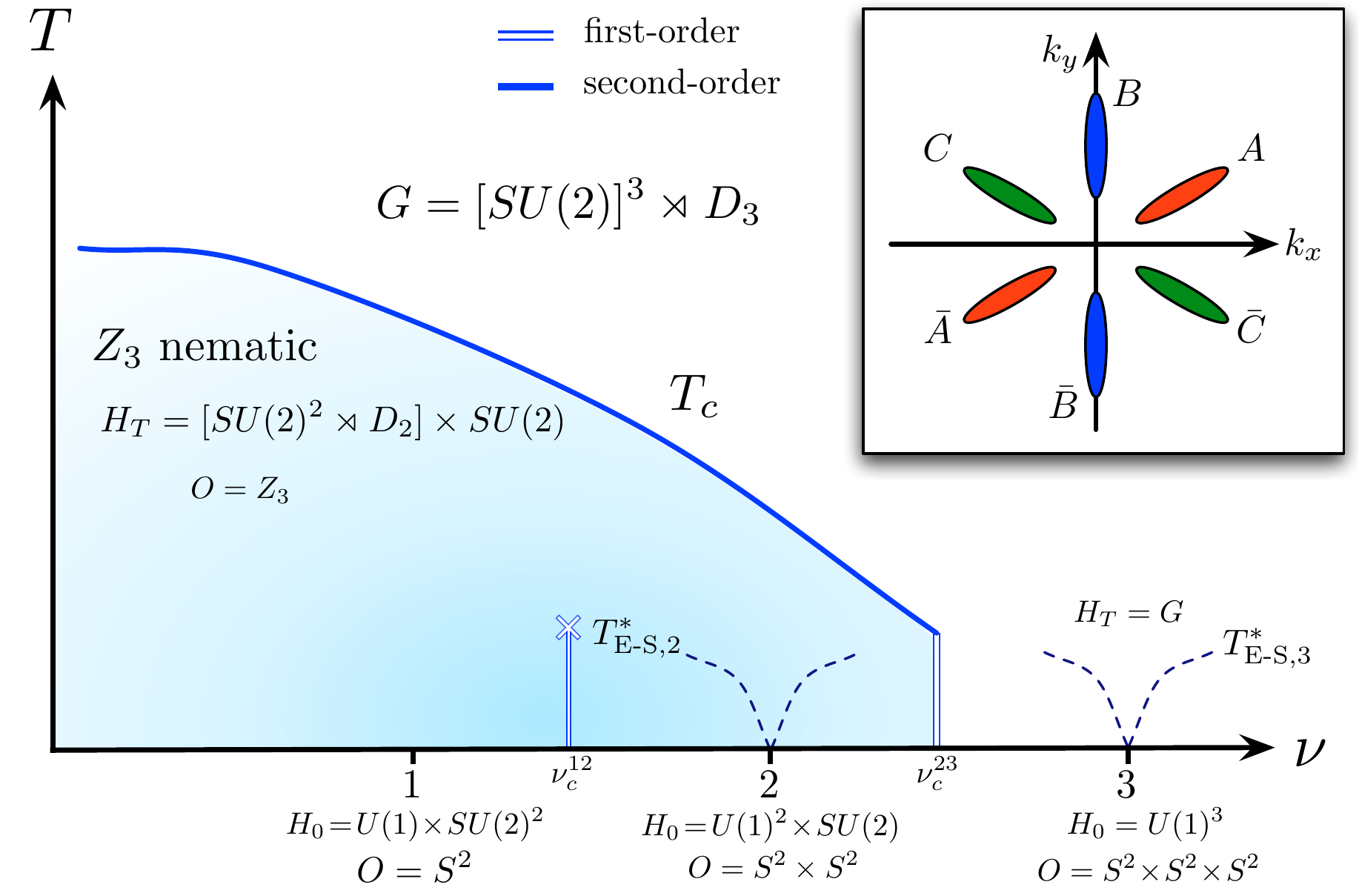}
\caption{{\bf Valley ordering in Si(111) QH states.} (Inset) Model Fermi surface. Ellipses denote constant-energy lines in $k$-space. (Main figure) Schematic global phase diagram, showing how the $G = [SU(2)]^3\rtimes D_3$ symmetry is broken to $H_0, H_T$ at zero and finite temperature. The order parameter spaces are $O = G/H_T$ for $T>0$, and $O = H_T/H_0$ at $T=0$. For $\nu=1,2$, ${D}_3$ symmetry breaks continuously at $T_c$, but this becomes first-order around $\nu=3$. Near $\nu=2,3$ order by doping yields to thermal order-by-disorder at $T\sim T^*_{\text{E-S}}$.} \label{fig:phasediag_bandstructure}
\end{figure}

In this paper, we revisit this interplay in the context of the QH states of multi-valley semiconductors, specifically those recently observed in two-dimensional electron gases (2DEGs) confined in Si(111) and Si(110) quantum wells~\cite{Ando}. 
We find several striking new phenomena embedded in surprisingly intricate phase diagrams --- even while considering only integer quantum
Hall states in the lowest Landau level (LLL). These systems exhibit 
six- and four-fold valley degeneracies in their respective electronic dispersions, as shown in Figures \ref{fig:phasediag_bandstructure} and \ref{fig:Si110}. Consequently, they exhibit large symmetry groups ---
$[SU(2)]^3\rtimes D_3$  for the (111) case 
and $[SU(2)]^2\rtimes D_2$  for the (110) example
--- in the standard limit where the magnetic length $\ell_B$ is much longer than the lattice constant $a$. Here, $D_n$ is the dihedral group of symmetries of a regular $n$-gon, and the semidirect product structure (denoted by $\rtimes$) reflects the fact that these discrete symmetries act upon the $SU(2)$ axes. The  rich phase structure derives from the various possibilities for breaking these symmetries, and how these manifest
at different $\nu$. For our primary example---the (111) system---we find a finite-temperature ${Z}_3$ transition into a nematic phase where the discrete factor is broken, and zero-temperature phases where the continuous $[SU(2)]^3$ symmetry is broken down to various subgroups. We sketch the phase diagram resulting from fitting together these possibilities in Fig.~\ref{fig:phasediag_bandstructure}.

The mechanisms of symmetry breaking are also unusual. While the
nematicity is driven
by  Hartree-Fock exchange interactions as is standard in QH ferromagnetism, the $T \rightarrow 0$ ordering  
involves  entropic selection --- {\it order by disorder}~\cite{villain_order_1980, moessner_low-temperature_1998, chalker_hidden_1992, chubukov_order_1992} --- and selection via the energetics of Skryme lattices that form
in the vicinity of integer $\nu$, a new mechanism that we term {\it order by doping}  in tribute to its entropic cousin. To our knowledge, this is the first time either selection mechanism has  been shown to operate in the QHE setting.

\noindent\textbf{Si(111):} We begin our discussion by listing some salient features of Si(111) quantum wells relevant to understanding the QHE in these systems. Since $g \approx 2$ in Si 
 we may assume the electrons to be spin polarized.  
  The valley degeneracy of the Si 2DEG depends on the orientation of the interface, as this choice can break the crystal symmetries responsible for the exact valley degeneracy in bulk Si. In case of the (111) interface, effective mass theory predicts a six-fold degeneracy~\cite{stern_properties_1967} (Fig.~\ref{fig:phasediag_bandstructure}, inset). This degeneracy is quite robust --- for instance, it cannot be lifted by changing the width of the confining well or by an interface potential. For the bulk of this paper, we take this degeneracy to be exact, surely an idealization; we comment on  corrections to this scenario at the end.

We label valleys as shown in Fig.~\ref{fig:phasediag_bandstructure} (inset). Valley $\kappa$ is centered at $\vec{K}_\kappa$, where 
 $\vec{K}_{A}=(\frac{\sqrt{3}K_0}{2},\frac{K_0}{2})$, $\vec{K}_{B}=(0,K_0)$ and $\vec{K}_{C}=(\frac{\sqrt{3}K_0}{2},-\frac{K_0}{2})$, with $\vec{K}_{\bar{\kappa}} = -\vec{K}_{\kappa}$. Here $K_0 \approx 1/a$ where $a$ is the lattice constant. Note that in each valley the effective mass tensor is anisotropic; 
 this is most evident in a coordinate system in which the mass tensor is diagonal. The single-particle Hamiltonian in valley $\kappa $ (where $\kappa=A,B,C$)  is 
  $H_{\kappa} = \sum_{i=1,2} \frac{( (\vec{p} + e\vec{A}/c - \vec{K}_{\kappa}) \cdot \vec{\eta}_{\kappa i})^2}{2 m_i}$, where $\vec{\eta}_{A/C1} = (\mp 1, \sqrt{3})$, $\vec{\eta}_{A/C2} = (\pm \sqrt{3}, 1)$, $\vec{\eta}_{B 1} = (1, 0)$ and $\vec{\eta}_{B 2} = (0, 1)$; $H_{\bar{\kappa}}$ is obtained by taking $K_0\rightarrow-K_0$ in these expressions.

In Landau gauge $\vec{A}=(0,Bx)$, the LLL eigenfunctions labeled by momentum $k_y$ are  given by
\begin{equation} \label{eq4}
\phi_{\kappa,k_y} = \frac{(f_{\kappa})^{1/4}}{(\pi^{1/2} \ell_B L_y)^{1/2}} e^{i \vec{K_{\kappa}}.\vec{r}} e^{i k_y y} e^{-(f_{\kappa} + i g_{\kappa})\frac{(x + k_y \ell_B^2)^2}{2 \ell_B^2}}, 
\end{equation}
where $(f,g)_{A,\bar{A}}=(\frac{4\sqrt{\lambda}}{\lambda+3},\frac{\sqrt{3}(1-\lambda)}{\lambda+3})$, $(f,g)_{B,\bar{B}}=(\frac{1}{\sqrt{\lambda}},0)$, $(f,g)_{C,\bar{C}}=(\frac{4\sqrt{\lambda}}{\lambda+3},\frac{-\sqrt{3}(1-\lambda)}{\lambda+3})$, $\lambda=(m_2 / m_1)\approx3.52$ and the magnetic length $\ell_B=\sqrt{\frac{\hbar c}{e B}}$. 
We focus on filling fractions $\nu < 6$ and ignore mixing between different Landau levels (LLs). 
 As is usual in QH ferromagnets, even if we restrict to (near-)integer filling, the exact degeneracy between the valley degrees of freedom at single-particle level is lifted by interactions, which select a ground state at each integer filling $\nu<6$, and in doing so break one or more symmetries spontaneously. The question of precisely how this happens is our focus in the remainder.

\noindent\textbf{Effective Hamiltonian:} Since we are working in a degenerate manifold of the electron kinetic energy --- quenched by the magnetic field --- the effective Hamiltonian is comprised solely of interaction terms, that inherit the kinetic anisotropies through their dependence on the single-particle LL eigenfunctions.
In  the limit $K_0 \ell_B\gg 1$, the electron-electron interaction term is  
\begin{equation} 
\label{eq:Ham}
H = \frac{1}{2S}\sum_{\vec{q},\kappa,\kappa^{'}}  V(\vec{q}) \rho_{\kappa\kappa}(\vec{q}) \rho_{\kappa^{'}\kappa^{'}}(-\vec{q})
\end{equation}
where $S=L_x L_y$ is the total area, $\rho_{\kappa\kappa}$ is the density operator within valley $\kappa$ projected to the LLL and $V(\vec{q})=\frac{2\pi e^2}{q}$ is the matrix element of the Coulomb interaction. [A static background is omitted from (\ref{eq:Ham}) for clarity.] 

The Hamiltonian (\ref{eq:Ham}) has an approximate $G = [SU(2)]^3\rtimes~D^3$ symmetry. To see this, note that $H$ is invariant under $SU(2)$ rotations between the two valleys $(\kappa\bar{\kappa})$ in the pair, explaining the $[SU(2)]^3$, as well as under a $D_6$ discrete point-group symmetry. However, any element of $D_6$ that only interchanges the two valleys ($\kappa,\bar{\kappa}$) in a pair is equivalent to an $SU(2)$ $\pi$-rotation; the $D_6$ elements not of this type form a $D_3$ subgroup that acts on the 3 $SU(2)$ indices, leading to the semidirect product structure~\cite{Supplement}. Recent work on wide (001)  AlAs quantum wells studied a symmetry similar to the discrete rotation above~\cite{abanin_nematic_2010, kumar_microscopic_2013}.

We  now consider the cases of various integer filling fractions using the vanishing mass anisotropy limit as the starting point. (In the theoretically convenient case of $\lambda=1$, $H$ is $SU(6)$ symmetric; qualitative pictures obtained for $|\lambda -1|\ll 1$ remain valid even when the anisotropy is no longer small~\cite{kumar_microscopic_2013}.) 
Note that, owing to particle-hole symmetry about $\nu=3$ 
the problems at $\nu=1,5$ are equivalent, as are those at $\nu=2,4$. This leaves us three distinct fillings to consider. 

\noindent\textbf{$\boldsymbol{\nu=}\mathbf{1, 5}$}: At the $SU(6)$ symmetric point $\lambda=1$,  the degenerate ground states with one filled LL are  $|\psi \rangle = \prod_{k_y} d_{k_y}^{\dagger} |0 \rangle$ where  $d_{k_y}^{\dagger} = \sum_{\kappa}\alpha_{\kappa} c^\dagger_{\kappa k_y}$, with $\sum_\kappa |\alpha_\kappa|^2 =1$
The anisotropy splits this degeneracy. 
At $|\lambda-1| \ll 1$, first-order perturbation theory yields
\begin{eqnarray}
\langle \psi |H| \psi \rangle & =& \!\!\!\!\!\!\sum_{\substack{\kappa,\kappa,'\sigma\\ \in\{A,B,C\}}}\!\!\!\!\!\! \delta V_{\kappa\kappa'}^\sigma 
(|\alpha_{\kappa}|^2 + |\alpha_{\bar{\kappa}}|^2)(|\alpha_{\kappa}|^2 + |\alpha_{\bar{\kappa}'}|^2)
\end{eqnarray}
where $\delta V_{\kappa\kappa'}^\sigma =\frac{1}{2}|\epsilon_{\sigma\kappa\kappa'}| (V_{\sigma\sigma}^{\sigma\sigma} - V_{\kappa\kappa'}^{\kappa'\kappa})$, and
\begin{eqnarray}
V_{\kappa \kappa'}^{ \kappa' \kappa}\!=\! \sum_{k,k'}\!\int_{\vec{r},\vec{r}{'}}\!\!  
 \phi_{\kappa k}^*(\vec{r}) \phi_{\kappa' k'}^*(\vec{r}\,{'})V_{\vec{r}-\vec{r}{'}} 
\phi_{\kappa' k}(\vec{r}\,{'}) \phi_{\kappa k'}(\vec{r}).
\end{eqnarray}
The 
$\delta V_{\kappa\kappa'}^\sigma$  
are all positive and proportional to $N$, the number of electrons. Hence the approximate new ground states in the thermodynamic limit are of the form $\ket{\kappa} = \prod_{k_y} (\alpha_{\kappa} c^{\dagger}_{\kappa,k_y} + \alpha_{\bar{\kappa}} c^{\dagger}_{\bar{\kappa},k_y}) \ket{0}$; 
 these break the $[SU(2)]^3\rtimes~D_3$ symmetry down to $H_0=U(1)\times[SU(2)^2\rtimes D_2]$ (where the second factor refers to rotations of the unoccupied pairs), leading to a single Goldstone mode (as $G/H_0 = S^2$). 
Working in the vicinity of $\nu=1$ at $T=0$, from standard energetic arguments~\cite{sondhi_skyrmions_1993} we conclude that for $\nu \gtrsim 1$, skyrmions are created within the occupied-valley subspace $(\kappa,\bar{\kappa})$; similarly, for $\nu \lesssim 1$, anti-skyrmions are created~\cite{sondhi_skyrmions_1993}. 
At any $T\neq0$, statistical averaging over Goldstone modes restores the broken $SU(2)$ symmetry, so that the invariance group is $H_T = SU(2)\times[SU(2)^2\rtimes D_2]$. The order parameter of the resulting phase lies in  $G/H_T= {Z}_3$~\cite{Supplement}. We conclude that 
valley ferromagnetic order onsets via a finite temperature  ${Z}_3$ transition into a nematic phase with broken orientational symmetry (Fig.~\ref{fig:phasediag_bandstructure}).

\begin{figure}[tb]
\centering
\includegraphics[width=\columnwidth]{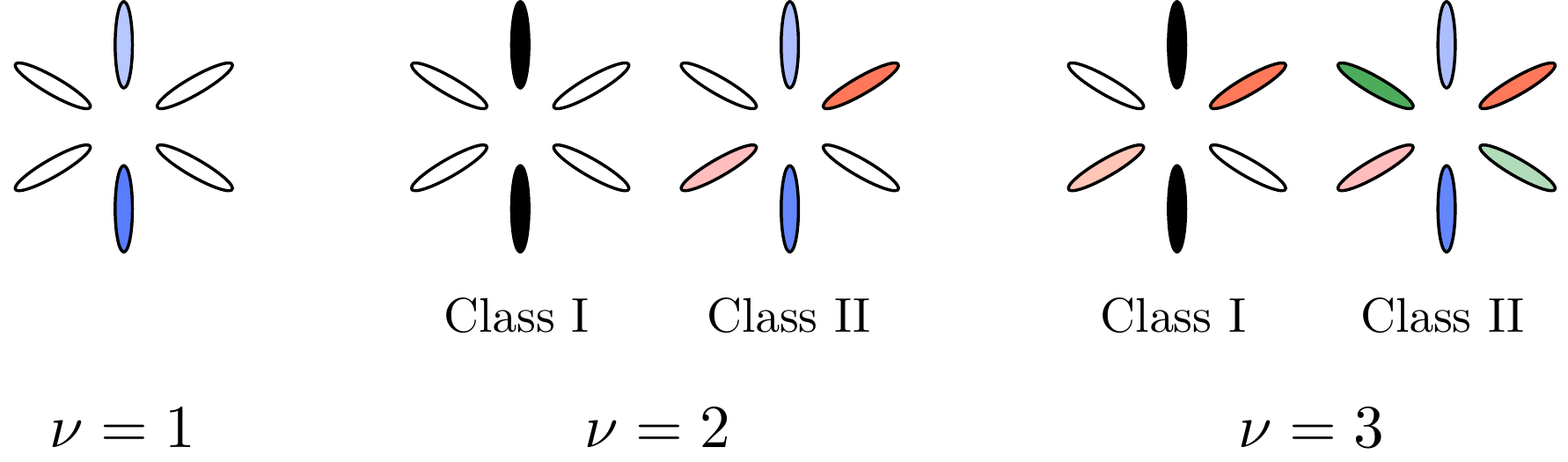}
\caption{{\bf Possible valley-ordered  states at $\nu=1,2,3$}, including representatives of Class I and II states for $\nu=2,3$. Unfilled and fully-filled valleys are shown as empty and filled ellipses; valleys partially-filled due to a particular choice of $SU(2)$ vector within the two-valley subspace are shaded.} 
\label{fig:Ordering}
\end{figure}

\noindent\textbf{$\boldsymbol{\nu=}\mathbf{2, 4}$}: We next consider the case when two LLs are filled. Again, we begin at  $\lambda=1$ where the degenerate ground states are given by $|\psi \rangle =  \prod_{i=1,2} \prod_{k_y} d_{i,k_y}^{\dagger} |0 \rangle$ where 
 $d_{i,k_y}^{\dagger} = \sum_{\kappa} \alpha_{\kappa i} c^\dagger_{\kappa k_y}$, and $\sum_{\kappa} |\alpha_{\kappa i}|^2=1$.
 Moving away from the $SU(6)$ point, but keeping $|\lambda-1| \ll 1$, the ground state manifold has two kinds of states that remain degenerate even upon inclusion of the anisotropic terms  (Fig. \ref{fig:Ordering}). ``Class I" ground states are obtained by filling {\it both} valleys in a pair, and take 
the form $\ket{\kappa\bar{\kappa}}= \prod_{k_y} c_{\kappa,k_y}^{\dagger} c_{\bar{\kappa},k_y}^{\dagger} |0 \rangle$. ``Class II" ground states  
on the other hand are constructed by picking two pairs and setting each to have $\nu=1$ by spontaneously breaking the residual $SU(2)$ symmetry of rotations within the pair. These are of the form $\ket{\kappa\kappa'} = \prod_{k_y} (\alpha_{\kappa} c^{\dagger}_{A,k_y} + \alpha_{\bar{\kappa}} c^{\dagger}_{D,k_y}) (\alpha_{\kappa'} c^{\dagger}_{\kappa',k_y} + \alpha_{\bar{\kappa}'} c^{\dagger}_{\bar{\kappa}',k_y})\ket{0}$. While Class I states break the $[SU(2)]^3\rtimes D_3$ symmetry down to $[SU(2)]^3$, Class II states break it down to $U(1)^2\times SU(2)$; the order parameter spaces are $Z_3$ and $S^2\times S^2$ for class I and II states, respectively~\cite{Supplement}, which therefore host no and two Goldstone modes. This disparity leads to selection of the latter by thermal fluctuations as we discuss below.
 
First, however, we demonstrate that selection occurs due to charge doping -- incommensuration -- at $T=0$. To  see why this is so, observe that  doping Class II states to a filling $\nu\gtrsim 2$ ($\nu\lesssim 2$) proceeds by creating skyrmions (anti-skyrmions) in the two-dimensional subspaces of the occupied valley pairs.  
For Class I states on the other hand, the charge added or subtracted is accomodated in a conventional quasielectron (quasihole) Wigner crystal. As  skyrmion (anti-skyrmion) lattices have lower energy~\cite{sondhi_skyrmions_1993}, we argue that doping selects Class II states.

Turning now to $T>0$, we observe  that  the combination of a high ground state degeneracy and a disconnected ground state manifold --- there is no continuous path in the set of ground states that connects a state in Class I to a state in Class II ---
 is ideal for seeing ``order by disorder". This phenomenon, in which entropic considerations select a ground state,  occurs often in frustrated spin systems~\cite{villain_order_1980, moessner_low-temperature_1998, chalker_hidden_1992, chubukov_order_1992}.  Since there are gapless excitations about Class II states, they are selected by thermal fluctuations as the free energy of fluctations about the degenerate ground state manifold is peaked about states with a large number of soft modes. However this mechanism comes into play above a crossover scale $T^*$ (Fig. \ref{fig:phasediag_bandstructure}). {\it Below }$T^*$, energetic, rather than entropic, considerations favor Class II states ---  the order by doping mechanism.  The crossover between selection by energetic considerations and selection by entropy --- that we dub the ``E-S crossover''--- takes place at $T^*_{\text{E-S}}  \sim \frac{e^2}{\ell_B}$. 

We observe that while Goldstone modes are responsible for order by disorder, the $SU(2)$ symmetries remain unbroken at any $T\neq 0$  so that the invariance group is $H_T = [SU(2)\rtimes D_2]\times SU(2)$, (the $D_2$ reappears as we may once again interchange between the filled pairs when $SU(2)$ is restored). As $G/H_T = Z_3$, we conclude that the system has a transition at $T_c>0$ described by a $Z_3$ nematic order parameter, in which the $D_3$ symmetry is broken (Fig. \ref{fig:phasediag_bandstructure}). Since in the experimental systems of interest the filling is tuned with field rather than by gating, we anticipate that $T_c^{\nu=2} < T_c^{\nu=1}$, as the relevant energy scale is the surface tension of $\mathbb{Z}_3$ domain walls, that depends in turn on the Coulomb energy $e^2/\ell_B \propto \sqrt{B}$ \footnote{There is also a factor of $1/2$ from the fact that at fixed electron number only half of the electrons enter the domain wall energetics at $\nu=2$}.

\noindent\textbf{$\boldsymbol{\nu=}\mathbf{3}$}: We now examine the situation with three filled LLs. At the isotropic point $\lambda=1$, the degenerate ground  states are $|\psi \rangle =  \prod_{i=1,2,3} \prod_{k_y} d_{i,k_y}^{\dagger} |0 \rangle$ where $d^\dagger_{i, k_y} =  \sum_{\kappa}\alpha_{i\kappa} d^\dagger_{\kappa k_y}$.
 As before, the symmetry is reduced for $\lambda\neq 1$, and the new ground state manifold again has two kinds of states  (Fig. \ref{fig:Ordering}). Class I states are obtained by filling both valleys in a pair and then spontaneously breaking  $SU(2)$ in another pair, and are of the form $\ket{\kappa\bar{\kappa}\kappa'} = \prod_{k_y} c_{\kappa,k_y}^{\dagger} c_{\bar{\kappa},k_y}^{\dagger} (\alpha c^{\dagger}_{\kappa',k_y} + \alpha' c^{\dagger}_{\bar{\kappa}',k_y})\ket{0}$. Class II ground states are obtained by breaking each of the three $SU(2)$s by forming a $\nu=1$ state in each pair, so $\ket{ABC} = \prod_{k_y} \prod_{\kappa=A,B,C}(\alpha_\kappa c^{\dagger}_{\kappa,k_y} + \alpha_{\bar{\kappa}} c^{\dagger}_{\bar{\kappa},k_y}) \ket{0}$ with $|\alpha_\kappa|^2 +|\alpha_{\bar{\kappa}}|^2 =1$. 
Class I states are invariant under $H_0 = SU(2)^2\times U(1)$, and Class II states under $U(1)^3$. By explicit construction~\cite{Supplement} we find the order parameter spaces $S^2$ and $S^2\times S^2\times S^2$, leading to one and three Goldstone modes respectively.
 
Consider charge doping at $T=0$. For $\nu\gtrsim 3$ ($\nu\lesssim 3$), skyrmions (anti-skyrmions) are created about both Class I and II states. However, the structure of the resulting triangular lattices is quite different. Discussing skyrmions for specificity, for Class I states doping proceeds  by making a lattice of skyrmions that live in only one two-dimensional subspace, whereas for Class II states the skyrmion lattice has a tripled unit cell, as it results from symmetrically combining three sublattices each built from skyrmions in one of the three different two-dimensional subspaces. As there is no valley Zeeman energy, the skyrmion size is set entirely by their density. In order to estimate the energies of the competing skyrmion crystals, we utilize the fact that the relevant nonlinear sigma models optimize their gradient energy for {\it analytic} (two-component)  spinor solutions -- non-analytic configurations generically have higher energy. The simplest such analytic solution that is (quasi-)periodic  with finite topological charge $Q$ per unit cell, has $Q=2$, as all $Q=1$ configurations with these desiderata are non-analytic and hence have higher energy~\cite{skyrme_lattice}. Such a quasi-periodic $Q=2$ spinor solution~\cite{multicomponent_roderich_2013}  gives Class I and II states identical gradient energies, so  the issue turns on the Coulomb energy which is lower for Class II. 
We therefore conclude that doping selects Class II states.

For $T>0$, the Goldstone mode fluctuations about Class II states restore the full $G = [SU(2)]^3\rtimes D_3$ symmetry and hence there is no sharp finite-temperature transition owing to the lack of any broken symmetries. We nevertheless expect thermal selection of Class II states owing to the excess of Goldstone modes compared to Class I states. As in the $\nu=2$ case this occurs above a scale $T^*_{\text{E-S}}$, below which  order by doping  dominates.  

In combining the results for $\nu=1, 2,3$ we note that their distinct symmetries and doping energetics at $T=0$ point to first-order transitions at $\nu^{12}_c$ and $\nu_c^{23}$, where $1<\nu^{12}_{c}<2<\nu_{c}^{23}<3$, that we expect survive to $T>0$. Thus,  we arrive at the global phase diagram of Fig.~\ref{fig:phasediag_bandstructure}.

\noindent\textbf{Si(110):} We briefly discuss how these ideas apply in modified form to 
 (110) oriented wide Si quantum wells. In the presence of a strong interface potential, effective mass theory predicts a fourfold valley degeneracy~\cite{stern_properties_1967} (Fig.~\ref{fig:Si110}), with valleys centered at  $\vec{K}_{A}=-\vec{K}_{\bar{A}}=(K,0)$ and $\vec{K}_{B}=-\vec{K}_{\bar{B}}=(0,K)$. Once again working in Landau gauge $\vec{A}=(0,Bx)$, the LLL eigenfunctions are given by Equation \ref{eq4} with $(f,g)_{A,\bar{A}}=(\sqrt{\lambda},0)$ and $(f,g)_{B,\bar{B}}=(\frac{1}{\sqrt{\lambda}},0)$. 
The interaction Hamiltonian has  $[SU(2)]^2\rtimes D_2$ symmetry
 where the $SU(2)$s are independent rotations in the pair subspaces  ($A,\bar{A}$) and ($B,\bar{B}$) and the $D_2$ symmetry interchanges these pairs~\cite{Supplement}. 
\begin{figure}[tb]
\centering
\includegraphics[width=\columnwidth]{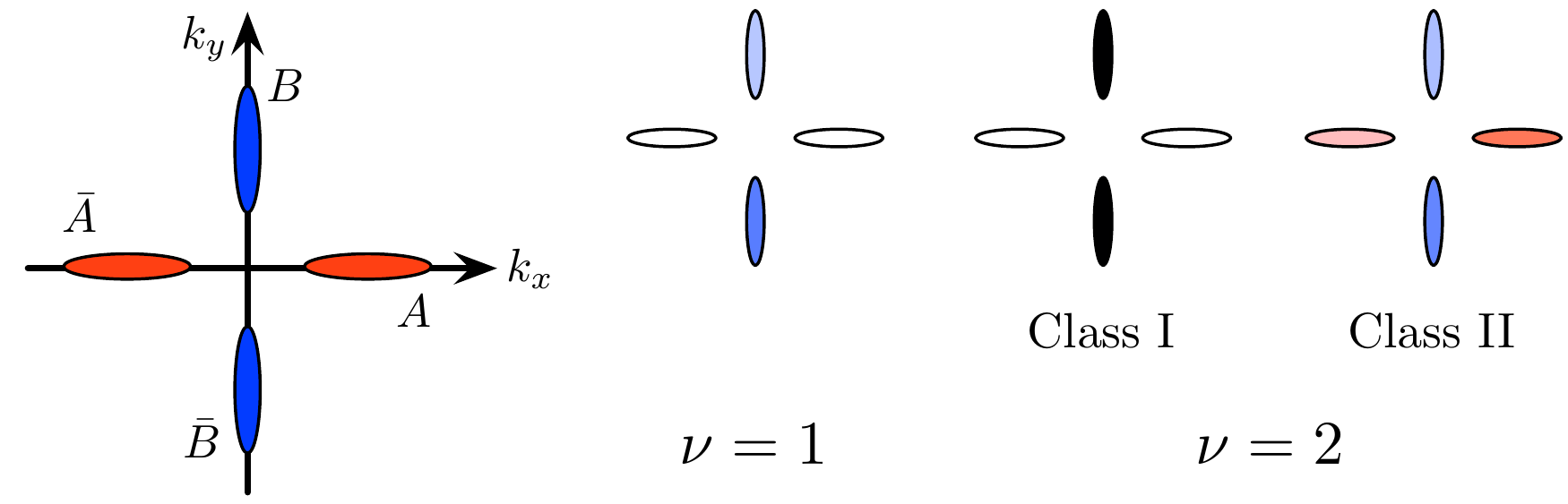}
\caption{{\bf Model Fermi surface and possible valley-ordered states for Si(110) quantum wells.}} \label{fig:Si110}
\end{figure}

For $|\lambda-1| \ll 1$, the ground states at $\nu=1$ are given by $|\psi \rangle =  \prod_{k_y} (\alpha_\kappa c^{\dagger}_{\kappa,k_y} + \alpha_{\bar{\kappa}}c^{\dagger}_{\bar{\kappa},k_y} \ket{0}$ and $|\alpha_\kappa|^2 + |\alpha_{\bar{\kappa}}|^2 = 1$ (Fig.~\ref{fig:Si110}). This case resembles  $\nu=1$ for Si(111) and we expect analogous results.

At $\nu=2$ the ground state manifold once again supports two types of states (Fig.~\ref{fig:Si110}). Class I states have the form $\ket{\kappa\bar{\kappa}} = \prod_{k_y} c_{\kappa,k_y}^{\dagger} c_{\bar{\kappa},k_y}^{\dagger} |0 \rangle$. Class II  states are of the form $\ket{AB}  = \prod_{k_y}\prod_{\kappa=A,B} (\alpha_\kappa c^{\dagger}_{A,k_y} + \alpha_{\bar{\kappa}} c^{\dagger}_{\bar{\kappa},k_y})\ket{0}$ where $|\alpha_\kappa|^2 +|\alpha_{\bar{\kappa}}|^2 =1$. Though the mode counting --- zero versus a pair of Goldstone modes --- is similar to  $\nu=2$  for Si(111), at $T=0$ the full $[SU(2)]^2\rtimes D_2$ symmetry is restored to Class II states by thermal averaging, analogous to the Si(111) $\nu=3$ Class II state. Therefore, we expect selection of Class II by thermal fluctuations and charge doping, but no finite-temperature transition about $\nu=2$ in Si(110).

\noindent\textbf{Experiments:} We expect that nematic order leads to measurable anisotropies in longitudinal conductivities $\sigma_{xx}, \sigma_{yy}$, though the orientation of the valleys with respect to the symmetry axes for Si(111) may present an added complication absent in Si(110), even for samples oriented along crystallographic axes. For $\nu=1,2$ in Si(111) and $\nu=1$ in Si(110) the anisotropy should show order-parameter onset behavior at $T_c$, typically a few kelvin at $B\approx 10$T for systems with comparable mass anisotropy and dielectric constant~\cite{abanin_nematic_2010}. Class II state selection at $\nu=2$ in both Si(111) and Si(110) will be reflected by the extreme sensitivity of the activation gap to strain-induced valley Zeeman splitting~\cite{abanin_nematic_2010} absent in Class I states that lack skyrmion excitations. The selection of Class II states at $\nu=3$ in Si(111) is challenging to detect as they lack nematic order, and Class I states also host skyrmions. However, restoration of orientational symmetry in going from $\nu=2$ to $\nu=3$ coupled with observation of the QHE would bolster this scenario.

\noindent\textbf{Concluding Remarks:} In closing we comment briefly on complications ignored in our discussion. Foremost among these is the neglect of various terms in (\ref{eq:Ham}) that while suppressed by $O((a/\ell_B)^2)$ relative to the dominant Coulomb energy scale could compete with the selection mechanisms discussed above. For $B\approx 10$T, we find that these terms split energy levels by a few  millikelvin, so that there is a large window of temperature where their neglect is justified. We note that competition between quantum selection by high-order effects and thermal order-by-disorder was recently studied in quantum magnets~\cite{2014arXiv1409.7070C}; similar situations may arise here once the neglected terms become significant.

Secondly, in a more realistic situation, the six-fold valley degeneracy can be lifted due to wafer miscut and  strain arising from lattice mismatch. While valley splitting due to the former mechanism is negligible compared to the cyclotron gap~\cite{mcfarland_multi-valley_2010}, the latter \emph{can} be more significant~\cite{shashkin_strongly_2007, tsui_observation_1979}. Although this problem has been largely solved by working with 2DEGs on a H-terminated Si(111) surface~\cite{kott_valley-degenerate_2014}, both mechanisms can still change the E-S crossover temperature $T^*_{\text{E-S}}$ and possibly even the Class of states that are selected below $T^*$.

Finally, spatial disorder induces a random field acting on the nematic order parameter, which is a relevant perturbation. An infinitesimal field destroys the nematic order  in the thermodynamic limit~\cite{binder_random-field_1983, imry_random-field_1975} at $\nu=1,2$ by proliferating domain walls but the QHE survives as long as disorder is sufficiently weak~\cite{abanin_nematic_2010}. The interplay of disorder with the novel selection mechanisms discussed above is likely intricate and worthy of further study.

\noindent\textbf{Acknowledgements.} We thank Rahul Nandkishore, Joseph Maciejko, Ashvin Vishwanath and Romain Vasseur for useful discussions, Tomasz Kott and Bruce Kane for discussing their experimental data, and Dmitry Abanin and Steven Kivelson for collaboration on related work. This work was supported by NSF Grant Numbers DMR 1006608, 1311781 and PHY-1005429 (AK and SLS) and UC Irvine start-up funds (SAP).
\bibliography{NematicValleys}
\clearpage
\begin{appendix}
\onecolumngrid

\centerline{\bf Supplementary Material}

\setcounter{equation}{0}

\section{GROUP-THEORETIC ANALYSIS OF SYMMETRY BREAKING}

We discuss the simpler four-valley case first as a warm-up, before moving to the six-valley example. In each case, we first discuss the high-temperature symmetry-group $G$, the finite-temperature invariance subgroup of the broken-symmetry states $H_T$, and finally its zero-temperature counterpart $H_0$. The nonlinear sigma models (NL$\sigma$M) governing the $T>0$ and $T=0$ transitions have order-parameter spaces given by the group manifolds $G/H_T$ and $H_T/H_0$, respectively. Valley indices are as described in the main text.
\subsection{Four-Valley Case}
We first show that high-temperature valley symmetry group is $[SU(2)]^2\rtimes \gD_2$, where the $\gD_2$ interchanges the two $SU(2)$ axes. To see this, we note that the valley Hamiltonian (after including the anisotropy terms) has the following symmetries (refer to Fig.~3 of the main text for valley labeling): two distinct  $SU(2)$ symmetries that each act within a valley pair $(A,\bar{A})$ and $(B,\bar{B})$, and the dihedral group of symmetries of the square, that we denote $D_4$. 
The full symmetry group $G$ is obtained by combining these symmetries. Clearly  $N = [SU(2)]^2$ is a subgroup of $G$. Turning to $D_4$, we recall that this is generated by two operations, a permutation $r = (AB\bar{A}\bar{B})$  and a swap $\rho = (A\bar{A})$, where we use conventional notation to describe the action of finite groups: for instance $(a_1\ldots a_m)(b_1\ldots b_n)$ denotes the cyclic permutations $a_1\rightarrow a_2\rightarrow  \ldots\rightarrow  a_m\rightarrow a_1$ and $b_1\rightarrow b_2\rightarrow \ldots\rightarrow b_n\rightarrow b_1$ with all other `letters' left invariant.  Now, it is clear that $r^4 = e$, the identity; also, we note that $r^2$ simply interchanges the two valleys in a pair, and is therefore equivalent to a $\pi$ rotation within each valley pair, {\it i.e.} $r^2 
\in N$. Therefore it follows that the coset $r^2N =  N =  N r^2$. A similar argument reveals that $\rho N = N = N \rho$, since $\rho$ corresponds to a $\pi$-rotation in the pair $(A\bar{A})$ coupled with an identity operation in the other valley pair. 
Finally, we observe that since $r$ preserves the valley pair structure, transforming valley indices by $r$, performing independent $SU(2)$ rotations within each pair of valleys, and undoing the index transformation, must be equivalent to a product of independent rotations within each pair, so that $r^{-1} N r = N$. Since $D_4 = \{e, r, r^2, r^3, \rho, r\rho, r^2\rho, r^3\rho\}$, we see that full list of cosets is $\{N, rN\}$.  Thus, (i) $N$ is a normal subgroup of $G$, $N\triangleleft G$ and (ii) $G = \{ N, rN\}$, so that $G/N \cong \gD_2$. We therefore conclude that $G = N\rtimes \gD_2 = [SU(2)]^2\rtimes \gD_2$, where in identifying the $\gD_2$ structure we used the fact  $r^2\sim e$  in the coset space since $r^2N = N = eN$.

 We now turn to the breaking of symmetries at different fillings. We will discuss the symmetry breaking in two stages: first, we will determine the residual symmetry group $H_T$ for $T>0$, where the Mermin-Wagner theorem precludes the breaking of continuous symmetries; the corresponding  NL$\sigma$M has target space $G/H_T$. Then, we will discuss how $H_T$ is further broken down to $H_0$ at $T=0$, described by a NL$\sigma$M with target space $H_T/H_0$.

 At $\nu=1$, 
and finite temperature we choose to fill a single valley pair while leaving the other unfilled. The invariant subgroup $H_T$ of the resulting state is $SU(2)^2$ (corresponding to rotating in the filled and unfilled pairs -- since for $T>0$ Goldstone modes lead to averaging over all possible superpositions of valleys within the filled pair), but the semidirect product structure does {\it not} survive as we can distinguish the filled and unfilled pairs and therefore their interchange is not a symmetry. The corresponding NL$\sigma$M target space is $G/H_T = ([SU(2)]^2\rtimes \gD_2)/ [SU(2)^2] = \gD_2 \cong Z_2$, consistent with our argument that the symmetry is broken via a finite-temperature Ising transition. As $T\rightarrow 0$ the Goldstone mode fluctuations responsible for the finite-temperature restoration of $SU(2)$ symmetry (of rotating between valleys $(\kappa,\bar{\kappa})$ in the filled pair) are suppressed, and this symmetry is broken by a specific choice of $SU(2)$ vector in the $(\kappa,\bar{\kappa})$ subspace. This leaves a residual $U(1)$ phase, but the $SU(2)$ symmetry between the unfilled valleys is still preserved, and therefore we have the residual symmetry group $H_0 = U(1)\times SU(2)$, giving  the  target space $H_T/H_0 = [SU(2)\times SU(2)]/[U(1)\times SU(2)] = SU(2)/U(1) = S^2$.

For Class II states at $\nu=2$, we fill both valley pairs to $\nu=1$, but as before the $SU(2)$ symmetry between the two valleys in each pair is restored at any finite temperature. As a consequence, we can still interchange $SU(2)$ axes in this case, so we have $H_T = G =  [SU(2)]^2\rtimes \gD_2$, and so there is no finite-temperature phase transition as $G/H_T$ is the trivial group. We may parameterize any state in this base space $\vec{z}\in H_T$ by $\vec{z} = g\cdot \vec{z}_0$, where 
\beq
g = \left( \begin{array}{cc} g_1 & 0 \\ 0 &g_2  \end{array}\right), 
\eeq
with $g_{1}, g_{2}\in SU(2)$, and $\vec{z}_0 = (1,0,1,0)^T$ is a reference spinor; note that this implicitly respects the semidirect product structure of $H_T$.  As $T\rightarrow 0$, we break each of the two $SU(2)$ symmetries down to $U(1)$. This remaining $U(1)$ invariance within each valley pair is generated by matrices $h \in H_0$, where
\beq
h = \left( \begin{array}{cc} h_1 & 0 \\ 0 &h_2  \end{array}\right), 
\eeq
 with $h_{1}, h_{2} \in U(1)$. Using the equivalence relation on $\vec{z},\vec{z}^{'}$ given by $\vec{z} \sim \vec{z}^{'} \iff \vec{z}^{'}=  h\cdot g'\vec{z}$ for $h\in H_0$, we see that $H_T/H_0\cong S^2\times S^2$.
For Class I states at $\nu=2$, we fill both valleys in a pair, and therefore the residual symmetry group is $H_T = SU(2)\times SU(2)$ as this corresponds to rotating within the filled and unfilled pairs -- we cannot interchange between the pairs. So, there is a putative finite-temperature transition possible for Class I states, as $G/H_T= \gD_2 \cong Z_2 $. However, as $T\rightarrow 0$, we observe that there is no additional structure that emerges as the full $SU(2)\times SU(2)$ symmetry is preserved. In contrast, Class II states have Goldstone modes that will lead to their selection over Class II states as $T\rightarrow 0$. Therefore starting from $T=0$ and restoring symmetry in stages we see that Class I states do not emerge in the phase diagram.

\subsection{Six-Valley Case}
In the six-valley case, we begin by observing that the high-temperature symmetry group is $G = [SU(2)]^3\rtimes \gD_3$, where $\gD_3$ is the dihedral group of symmetries of an equilateral triangle (isomorphic to $S_3$, the symmetric group) that acts on the three $SU(2)$ axes. To see this, we first observe that the symmetries of the valley Hamiltonian are (i) three $SU(2)$ symmetries that rotate between the two valleys in each pair $(A\bar{A}), (B\bar{B}), (C\bar{C})$ and (ii) the dihedral group $D_6$ of symmetries of a regular hexagon. Clearly, $N = SU(2)^3$ is a subgroup. Turning next to $D_6$, we observe that it is generated by the sixfold rotation $r = (ABC\bar{A}\bar{B}\bar{C})$ and the reflection $\rho = (A\bar{A})(BC)(\bar{B}\bar{C})$, so that we may write
 $D_6 = \{e, r, r^2, r^3, r^4, r^5, \rho, \rho r, \rho r^2, \rho r^3, \rho r^4, \rho r^5\}$. [Note that, {\it unlike} the swap in the four-valley case, here we cannot write $\rho$ as equivalent to a rotation; while it is indeed a rotation in the indices $(A\bar{A})$, it is {\it not} on the remaining indices.] 
 In listing cosets, we first observe that left and right cosets must be equivalent, {\it i.e.} $s^{-1}Ns = N$ for any $s \in D_6$, following the same logic as in the four-valley case: performing a discrete transformation $s$ on the valley indices, performing independent $SU(2)$ rotations in each valley pair and then transforming back to the original valley indices using $s^{-1}$ should be simply equivalent to three independent $SU(2)$ rotations, as long as $s$ preserves the pairing of the valleys, and it is clear this is satisfied by $r$, $\rho$, and hence by any combination of their powers. From this reasoning, we conclude that $N$ is a normal subgroup. In addition, we note that $\rho^2=e$, and furthermore $r^3 = (A\bar{A}) (B\bar{B})(C\bar{C})$, corresponding to $\pi$-rotations in each valley pair, whence $r^3N = N$. Putting these arguments together, we find the list of cosets to be $\{N, rN, r^2 N, \rho N,\rho rN, \rho r^2N\}$. Now, the fact that left and right cosets are equivalent means that the coset space has a group structure, obtained by simply writing $g_1 N g_2N =  g_1 g_2 N$. We can verify that the operations $r,\rho$ satisfy the identity $\rho^{-1} r^{-1} \rho r =  r^2$, so that under the coset multiplication rule $(r\rho)^{-1} \rho rN = r^2N \neq N$ {\it i.e.} the quotient group $G/N$ is non-Abelian. Since the unique non-Abelian group of order $6$ is $\gD_3$, we see that $G/N \cong S^3$, from which the structure $G = [SU(2)^3]\rtimes \gD_3$ follows immediately.

We now discuss symmetry breaking using the same conventions as previously. 
 At $\nu=1$ and $T>0$, we fill a single valley, breaking the $\gD_3$ structure, but thermal fluctuations of Goldstone modes restore the $SU(2)$ symmetry within the filled pair. We still have the ability to perform $SU(2)$ rotations within the unfilled pairs as well as swap their axes; we argue that this leads to an $SU(2)^2\rtimes \gD_2$ structure within the unfilled subspace, so that in total we have $H_T \cong SU(2)\times[SU(2)^2\rtimes \gD_2]$. From this, we see that the NL$\sigma$M target space is given by $G/H_T = [SU(2)^3\rtimes \gD_3]/[SU(2)\times[SU(2)^2\rtimes \gD_2]] \cong Z_3$. At  $T=0$, we argue in analogy with the 4-valley case that $SU(2)$ in the filled pair is broken down to $U(1)$ so that the  residual symmetry  $H_0 = U(1)\times [SU(2)^2 \rtimes \gD_2]$, so that the target space is $H_T/H_0 \cong S^2$.
 
Turning now to $\nu=2$ and $T>0$, we first consider class II states where we fill a pair of valleys breaking the $S^3$ structure but preserving $SU(2)$ via thermal restoration of symmetry, so that $H_T =[SU(2)^2\rtimes \gD_2]\times SU(2)$ again (but  the role of the filled and unfilled valleys are interchanged), and once again $G/H_T \cong Z_3$. At $T=0$, the situation is similar to $\nu=2$ for the four-valley case: the residual symmetry is $U(1)^2\times SU(2)$, yielding the target space $H_T/H_0 \cong S^2\times S^2$. For Class I states at $\nu=2$, we fill both valleys in a pair, leading to residual symmetry group $H_T = SU(2)\times [SU(2)^2\rtimes \gD_2]$; since $G/H_T \cong Z_3$, it appears that there an alternative finite-temperature transition. However, proceeding to $T=0$, we see that there is no additional symmetry breaking, i.e. $H_0 = H_T$, leading to a lack of Goldstone modes, and therefore near $T=0$ Class II states are selected. Since the thermal fluctuations of the Goldstone modes about class II states restores $[SU(2)^2\rtimes \gD_2]\times SU(2)$ symmetry, which returns to to the full symmetry $G$ via a finite temperature transition, we never access Class I states.

At $\nu=3$, for Class II states we fill each of the three valley pairs to $\nu=1$, and and $T>0$ we have $H_T = G = [SU(2)]^3\rtimes S^3$; once again $G/H_T$ is trivial, reflecting the fact that there is no finite-temperature transition. To examine further symmetry breaking we may parametrize this new base space $H_T$ similarly to the four-valley case: we write any spinor in this space as $\vec{z} = g\vec{z_0}$ with 
\beq
g = \left( \begin{array}{ccc} g_1 & 0 & 0 \\ 0 &g_2 &0 \\ 0& 0 &g_3 \end{array}\right), 
\eeq
with $g_{1}, g_{2}, g_{3}\in SU(2)$, and $\vec{z}_0 = (1,0,1,0,1,0)^T$ is a reference spinor. 
As $T\rightarrow 0$, each $SU(2)$ breaks to $U(1)$, so once again we consider states $\vec{z}, \vec{z}^{'}$ that are connected by a product of $U(1)$ rotations in each valley to be equivalent, yielding target space $H_T/H_0 \cong S^2\times S^2\times S^2$. For Class I states, we fill both members of one valley pair to $\nu=1$, while filling one of the other pairs at $\nu=1$, and leaving the third pair unfilled. The corresponding symmetry is simply $H_T = SU(2)^3$, and therefore $G/H_T \cong \gD_3$, suggesting a potential finite-temperature transition. However, as $T\rightarrow 0$, we break the $SU(2)$ of the filled pair down to $U(1)$, so that $H_0 =U(1)\times SU(2)^2$, and $H_T/H_0\cong S^2$; since the corresponding theory has one Goldstone mode compared to three about Class II states, $T=0$ state selection favors Class II states, and therefore we do not see Class I states as thermal fluctuations for $T\neq 0$ restore the full symmetry $G$  immediately.

\end{appendix}
\end{document}